\documentclass{emulateapj}

\usepackage{natbib}
\usepackage{amsmath}

\newcommand{\about}{$\sim\!\!$~}
\newcommand{\kms}{\,km\,s$^{-1}$}

\def\lsim{\hbox{\rlap{\raise 0.425ex\hbox{$<$}}\lower 0.65ex\hbox{$\sim$}}}
\def\gsim{\hbox{\rlap{\raise 0.425ex\hbox{$>$}}\lower 0.65ex\hbox{$\sim$}}}

\def\arcsec{\hbox{$^{\prime\prime}$}}
\def\arcdeg{\mbox{$^\circ$}}

\newcommand{\mean}[1]{\left \langle #1 \right \rangle}

\newcommand{\vsi}{\protect\hbox{$v_{\rm Si~II}$}}
\newcommand{\vsiz}{\protect\hbox{$v_{\rm Si~II}^{0}$}}

\shorttitle{}
\shortauthors{Foley et~al.}

\begin{document}

 \title{Linking Type I\lowercase{a} Supernova Progenitors and their Resulting Explosions}

\def\cfa{1}
\def\car{2}
\def\weiz{3}
\def\chile{4}
\def\lco{5}
\def\ut{6}
\def\mpg{7}
\def\clay{8}
\def\mine{9}

\author{
{Ryan~J.~Foley}\altaffilmark{\cfa,\clay},
{Joshua~D.~Simon}\altaffilmark{\car},
{Christopher~R.~Burns}\altaffilmark{\car},
{Avishay~Gal-Yam}\altaffilmark{\weiz},
{Mario~Hamuy}\altaffilmark{\chile},
{Robert~P.~Kirshner}\altaffilmark{\cfa},
{Nidia~I.~Morrell}\altaffilmark{\lco},
{Mark~M.~Phillips}\altaffilmark{\lco},
{Gregory~A.~Shields}\altaffilmark{\ut},
{Assaf~Sternberg}\altaffilmark{\mpg,\mine}
}

\altaffiltext{\cfa}{
Harvard-Smithsonian Center for Astrophysics,
60 Garden Street, 
Cambridge, MA 02138 USA
}
\altaffiltext{\car}{
Observatories of the Carnegie Institution for Science,
813 Santa Barbara Street,
Pasadena, CA 91101, USA
}
\altaffiltext{\weiz}{
Benoziyo Center for Astrophysics,
Faculty of Physics,
Weizmann Institute of Science,
Rehovot 76100, Israel
}
\altaffiltext{\chile}{
Departamento de Astronom\'{i}a,
Universidad de Chile,
Casilla 36-D,
Santiago, Chile
}
\altaffiltext{\lco}{
Las Campanas Observatory,
Carnegie Observatories,
Casilla 601,
La Serena, Chile 
}
\altaffiltext{\ut}{
Department of Astronomy,
University of Texas,
Austin, TX 78712 USA
}
\altaffiltext{\mpg}{
Max-Planck-Institut f\"{u}r Astrophysik,
Karl-Schwarzschild-Stra\ss e 1,
85741 Garching, Germany
}
\altaffiltext{\clay}{
Clay Fellow. Electronic address rfoley@cfa.harvard.edu .
}
\altaffiltext{\mine}{
Minerva Fellow
}

\begin{abstract}
Comparing the ejecta velocities at maximum brightness and narrow
circumstellar/interstellar Na~D absorption line profiles of a sample
of 23 Type Ia supernovae (SNe~Ia), we determine that the properties of
SN~Ia progenitor systems and explosions are intimately connected.  As
demonstrated by \citet{Sternberg11}, half of all SNe~Ia with
detectable Na~D absorption at the host-galaxy redshift in
high-resolution spectroscopy have Na~D line profiles with significant
blueshifted absorption relative to the strongest absorption component,
which indicates that a large fraction of SN~Ia progenitor systems have
strong outflows.  In this study, we find that SNe~Ia with blueshifted
circumstellar/interstellar absorption systematically have higher
ejecta velocities and redder colors at maximum brightness relative to
the rest of the SN~Ia population.  This result is robust at a
98.9--99.8\% confidence level, providing the first link between the
progenitor systems and properties of the explosion.  This finding is
further evidence that the outflow scenario is the correct interpretation
of the blueshifted Na~D absorption, adding additional confirmation
that some SNe~Ia are produced from a single-degenerate progenitor
channel.  An additional implication is that either SN~Ia progenitor
systems have highly asymmetric outflows that are also aligned with the SN
explosion or SNe~Ia come from a variety of progenitor systems where
SNe~Ia from systems with strong outflows tend to have more kinetic energy
per unit mass than those from systems with weak or no outflows.
\end{abstract}


\defcitealias{Foley11:vgrad}{FSK11}
\defcitealias{Sternberg11}{S11}


\section{Introduction}\label{s:intro}

Type Ia supernovae (SNe~Ia) appear to come from carbon-oxygen white
dwarfs (WDs) in a binary system \citep{Hoyle60}; however, it is still
unknown if they result from the merger of two WDs \citep[the
``double-degenerate'' scenario;][]{Iben84, Webbink84} or a single WD
accreting material from a non-degenerate companion \citep[the
``single-degenerate'' scenario;][]{Whelan73, Iben96}.
\citet{Howell11} and \citet{Wang12:prog} present recent reviews on the
subject.  With the exception of the recently discovered, nearby
SN~Ia~2011fe \citep{Li11:11fe, Nugent11, Bloom12, Brown12, Chomiuk12,
Margutti12}, there are only weak constraints on the nature of the
binary companion for individual SNe~Ia \citep[e.g.,][]{Maoz08,
Foley10:08ge, Foley12:09ig, Edwards12, Schaefer12}, and even then, it
is difficult to completely exclude scenarios for subsamples of SNe~Ia.
Similarly, there are several different models for SN~Ia explosions
\citep[e.g.,][]{Nomoto84:w7, Woosley86, Khokhlov91, Hillebrandt00,
Gamezo05, Pakmor10, vanKerkwijk10, Sim10} and limited observational
constraints \citep{Blondin11:2D}.  This is particularly troubling
since measurements of SNe~Ia were used to discover the accelerating
expansion of the Universe \citep{Riess98:Lambda, Perlmutter99} and
continue to be exquisite tools for constraining cosmological
parameters \citep{Wood-Vasey07, Riess07, Freedman09, Hicken09:de,
Kessler09:cosmo, Amanullah10, Conley11, Suzuki12}.

The recent observations of time variable, narrow ($\Delta v <
100$~\kms; resulting from either circumstellar or interstellar
material in the host galaxy) Na~D absorption in several SNe~Ia
\citep[SNe~2006X, 1999cl, and 2007le;][]{Patat07:06x, Blondin09,
Simon09} are perhaps the best evidence that at least some SN~Ia
progenitor systems have strong outflows --- the hallmark of certain
classes of single-degenerate progenitor systems.  In the two cases
with high-resolution spectroscopy, the variable component of the Na~D
absorption profile was blueshifted relative to the strongest component
of the profile.  Using a large sample of SNe~Ia with single-epoch
high-resolution spectroscopy, \citet[hereafter, S11]{Sternberg11}
determined that about half of all SNe~Ia with Na~D absorption have
blueshifted Na~D absorption relative to the position of the strongest
absorption (12/22 SNe~Ia).  In the same sample, only a quarter of
SNe~Ia (5/22) have redshifted absorption, while the remaining quarter
(5/22) have single or symmetric absorption profiles.  If the
absorption were the result of only interstellar gas, we would expect
the fraction of blueshifted and redshifted profiles to be equal.
\citetalias{Sternberg11} concluded that the difference in these
numbers was likely the result of a large fraction of SNe~Ia ($\gtrsim
25\%$ of SNe~Ia with detected Na~D absorption) coming from progenitor
systems with strong outflows.

The first SN~Ia with identified variable Na~D was SN~2006X
\citep{Patat07:06x}.  This SN was also exceptional in its high ejecta
velocity and high reddening \citep{Wang08:06x}.  With only the
detection of variable Na~D in SN~2006X and the non-detection of
variable Na~D in the low-velocity SN~2007af, \citet{Simon07}
hypothesized that variable Na~D was exclusive to SNe~Ia with high
ejecta velocities.  Since then, detections of variable Na~D in
SNe~1999cl \citep{Blondin09} and 2007le \citep{Simon09} have been made
--- both SNe~Ia with relatively high \ion{Si}{2} $\lambda 6355$
velocities, \vsi, near maximum brightness.

Additionally, ejecta velocity provides insight into the explosion.
Directly, the ejecta velocity (up to viewing angle effects of
asymmetric explosions) probes the kinetic energy per unit mass.  If
one assumes that all SNe~Ia have approximately the same total mass
\citep[e.g.,][]{Mazzali07} and relatively small asymmetries
\citep[e.g.,][]{Leonard05}, then ejecta velocity is a good
tracer of the kinetic energy of the explosion.

The strong correlation seen between the velocity gradient of the
\ion{Si}{2} $\lambda 6355$ feature near maximum brightness of a SN~Ia
and the relative redshifting of forbidden lines in its nebular
spectrum \citep{Maeda10:asym} places strong constraints on explosion
models.  The early-time velocity measures the characteristics of the
outer layers of the SN, while the late-time velocity measures the
characteristics of the inner ejecta.  The connection can be
interpreted as the result of an asymmetric explosion, where different
viewing angles result in different velocities at early and late times
\citep{Maeda10:asym}.  This asymmetric explosion interpretation can
further explain the relation between SN~Ia ejecta velocity (and
velocity gradient) and intrinsic color \citep{Foley11:vel, Maeda11}.

While ejecta velocity probes the characteristics of the SN explosion,
the line profile of narrow Na~D is influenced by the progenitor
environment.  There has been tenuous evidence that SNe~Ia with
variable Na~D tend to have higher-velocity ejecta.  To determine if
this trend is happenstance, we use the large \citet*[hereafter,
FSK11]{Foley11:vgrad} collection of maximum-brightness \ion{Si}{2}
$\lambda 6355$ velocities for SNe~Ia, \vsiz, supplemented by
additional data presented here to both expand the
\citetalias{Sternberg11} sample and assign ejecta velocities to that
expanded sample.

The rest of this manuscript is structured as follows.  In
Section~\ref{s:data}, we present new data to expand the
\citetalias{Foley11:vgrad} and \citetalias{Sternberg11} samples and
then match the expanded samples.  In Section~\ref{s:results}, we
examine the properties of the matched sample, finding that SNe~Ia with
blueshifted Na~D line profiles tend to have higher ejecta velocities
and redder colors than typical SNe~Ia.  We discuss this result and
conclude in Section~\ref{s:conc}.


\section{Data}\label{s:data}

In this section, we construct our sample.  We begin with the
\citetalias{Sternberg11} sample of SNe~Ia and add to it additional
SNe~Ia with new high-resolution spectroscopy.  We discuss the
\citetalias{Foley11:vgrad} sample of SNe~Ia, and add SNe~Ia with new
photometry and low-resolution spectroscopy to the sampl.  We exclude
all SNe~Ia with $\Delta m_{15} (B) > 1.5$~mag since we cannot
determine an accurate \vsiz\ for these SNe.  Our final sample, the F12
sample, is comprised of all spectroscopically normal SNe~Ia with
high-resolution spectroscopy and a measurement of \vsiz.

\subsection{High-Resolution Spectroscopy}

\citetalias{Sternberg11} present a sample of 35 SNe~Ia with
high-resolution spectroscopy primarily obtained with the Keck,
Magellan, and VLT telescopes.  For all spectra with any narrow Na~D
absorption at the recession velocity of the host galaxy, a zero
velocity for the system was assigned to the strongest Na~D component.
The SNe were then classified as having blueshifted, redshifted,
single, or symmetric profiles.  If, for example, the Na~D profile had
additional absorption offset primarily to the blue of the
zero-velocity component, it was classified as Blueshifted.  Depending
on the relative strengths of circumstellar and interstellar
absorption, it is possible in principle that some SNe~Ia with
strong-outflow progenitor systems could appear as Redshifted, Single,
or (possibly) Symmetric.  However, in a large sample of SNe with
circumstellar material (CSM), the Blueshifted objects should outnumber
objects in the Redshifted sample.  Of the 35 SNe~Ia,
\citetalias{Sternberg11} find that 12, 5, and 5 SNe~Ia have
blueshifted, redshifted, and single or symmetric narrow Na~D
absorption profiles, respectively, while 13 have no detected Na~D
absorption.

Using additional data, we modify the \citetalias{Sternberg11} sample.
Specifically, we exclude SN~2008ge (which had no Na~D absorption)
because of its known peculiarities \citep{Foley10:08ge}.
\citetalias{Sternberg11} decided not to include some older SNe~Ia in
their sample because of a possible publication bias (i.e., SNe without
detected or extraordinary Na~D line profiles may not have ever been
published).  Since we are comparing velocities within our sample, this
publication bias should not affect us significantly.  We therefore add
SN~1986G \citep{Rich87, DOdorico89}, which has a blueshifted Na~D line
profile, and SN~2001el \citep{Sollerman05}, which has a redshifted
Na~D line profile.  Although SN~2000cx has high-resolution spectra
\citep[showing no Na~D absorption;][]{Patat07:00cx}, we do not add
this SN to the \citetalias{Sternberg11} sample because of its peculiar
nature \citep{Li01:00cx}.

We also consider SN~1994D \citep{Ho95, King95, Patat96}.  Its Na~D
line profile is complicated by its very low recession velocity.  There
are three groups of absorption lines at \about 0, 200--250, and
700~\kms, respectively.  The feature at 700~\kms\ is clearly a single
absorption line.  If one interprets the features at \about
200--250~\kms\ as being associated with the Milky Way, then SN~1994D
would be classified as a Single object.  However, if those features
are associated with the SN or its host galaxy, then it would be
classified as Blueshifted.  In the \citetalias{Sternberg11} sample,
there were no features beyond \about 150~\kms\ relative to the
strongest absorption feature, so a \about 450--500~\kms\ offset would
be an extreme outlier.  Nonetheless, we do not claim to fully
understand the characteristics of potential SN~Ia progenitor outflows.
We consider the possibility of SN~1994D being a Blueshifted SN as low,
and for our main analysis we classify it as Single.  However, we also
performed the analysis with SN~1994D classified as Blueshifted.  The
main results are not significantly affected by this choice.

We add new high-resolution spectra of six SNe~Ia obtained with the
High-Resolution Spectrograph \citep[HRS;][]{Tull98} on the
Hobby-Eberly Telescope (HET) in 2001 and 2002.  The HRS observations
used the 600 $\ell$/mm grating centered at 6302~\AA, providing
wavelength coverage from 5300 to 7300~\AA\ at $R = 30000$, and were
made through a 3\arcsec\ diameter fiber.

\begin{figure*}
\begin{center}
\epsscale{0.58}
\rotatebox{90}{
\plotone{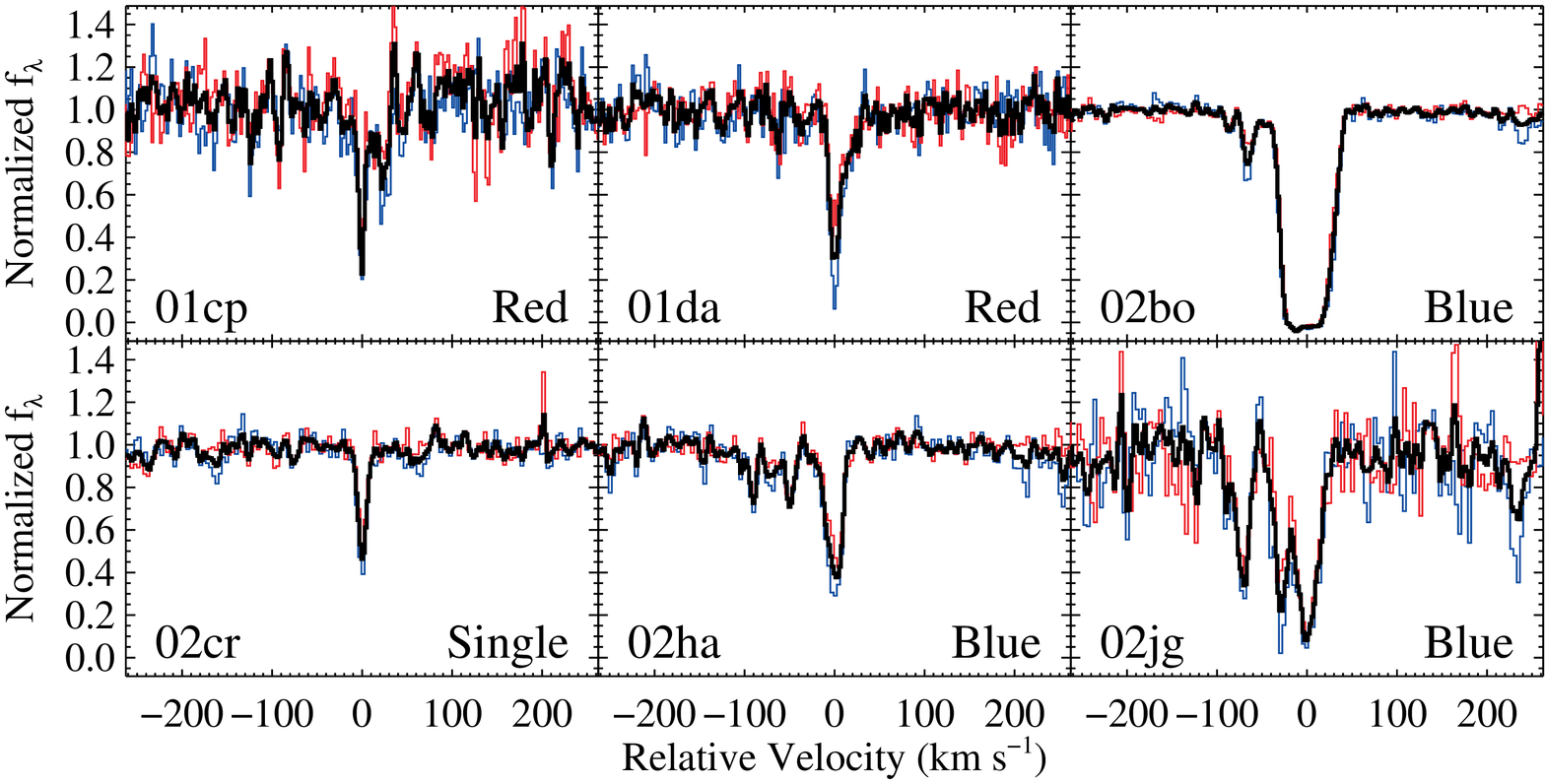}}
\caption{High-resolution HET spectra of SNe~2001cp, 2001da, 2002bo,
2002cr, 2002ha, and 2002jg.  The spectra are presented in normalized
flux units and only the region near Na~D is shown.  The data are
presented on a velocity scale where the zero velocity corresponds to
the strongest absorption.  The red and blue spectra are the individual
Na~D1 and Na~D2 components, respectively, while the black spectra
represent the combination of the two components.  We label each SN and
its profile designation in each panel.}\label{f:spec}
\end{center}
\end{figure*}

The HET data were reduced in IRAF with the {\sc echelle} package using
standard procedures.  Wavelength calibration was performed using
spectra of a ThAr comparison lamp obtained immediately before or after
the SN observations.  The Na~D emission lines in the HRS flatfield
lamps were masked out before flatfielding.  Emission from the night
sky was removed by subtracting the spectrum obtained through a sky
fiber after scaling to match the sky emission line intensities in the
SN spectrum (this scaling matches the relative throughputs of the
science and sky fibers derived from internal flatfield frames when
available).  The normalized-flux spectra near the host-galaxy Na~D
doublet are presented in Figure~\ref{f:spec}.

We report the properties of the expanded high-resolution SN~Ia sample
in Table~\ref{t:highres}.

\begin{deluxetable*}{llrcccc}
\tabletypesize{\scriptsize}
\tablewidth{0pt}
\tablecaption{Properties of the Expanded High-Resolution SN~Ia Sample\label{t:highres}}
\tablehead{
\colhead{} &
\colhead{} &
\colhead{} &
\colhead{$\Delta m_{15}(B)$} &
\colhead{$B_{\rm max} - V_{\rm max}$} &
\colhead{\vsiz} &
\colhead{Host} \\
\colhead{SN} &
\colhead{UT Date} &
\colhead{Phase\tablenotemark{a}} &
\colhead{(mag)} &
\colhead{(mag)} &
\colhead{($10^{3}$ \kms)} &
\colhead{Type}}

\startdata

\hline
\multicolumn{6}{c}{Blueshifted} \\
\hline
1986G           & 1986 May   9.4\tablenotemark{b}  & $-1.9$ (0.1) & 1.76 (0.01) & \phantom{$-$}0.88  (0.09) &           \nodata       & S0 pec \\
2002bo          & 2002 Mar.\ 21.3                  & $-2.4$ (0.7) & 1.24 (0.05) & \phantom{$-$}0.40  (0.04) &           $-13.1$ (0.2) & Sa \\
2002ha          & 2002 Nov.\ 2.6\tablenotemark{c}  &   0.2  (0.7) & 1.35 (0.05) &            $-0.09$ (0.03) &           $-11.3$ (0.2) & Sab \\
2002jg          & 2002 Nov.\ 30.1\tablenotemark{d} & $-1.4$ (0.1) & 1.54 (0.04) & \phantom{$-$}0.61  (0.03) &           \nodata       & Pair \\
2006X           & 2006 Feb.\ 18\tablenotemark{e}   & $-1.6$ (0.8) & 1.18 (0.03) & \phantom{$-$}1.26  (0.05) &           $-16.0$ (0.2) & Sbc \\
2007le          & 2007 Oct.\ 20.4                  & $-5.4$ (0.4) & 1.05 (0.04) & \phantom{$-$}0.29  (0.04) &           $-13.3$ (0.2) & Sc \\
2008C           & 2008 Jan.\ 17.4                  &  16.0  (0.3) & 1.07 (0.03) & \phantom{$-$}0.24  (0.03) &           $-11.3$ (0.2) & S0/a \\
2008dt          & 2008 Jul.\ 6.5                   &   6.5  (3.1) & 1.02 (0.14) & \phantom{$-$}0.55  (0.08) &           $-14.3$ (0.3) & S0/a \\
2008fp          & 2008 Sep.\ 17.3                  & $-4.1$ (0.1) & 0.92 (0.01) & \phantom{$-$}0.49  (0.01) &           $-10.7$ (0.2) & S0 pec \\
2009ig          & 2009 Oct.\ 16.3                  &  39.9  (0.1) & 0.89 (0.02) & \phantom{$-$}0.14  (0.04) &           $-13.5$ (0.2) & Sa \\
\hline
\multicolumn{6}{c}{Redshifted} \\
\hline
2001cp          & 2001 Jun.\ 28.3                  &   0.1  (0.4) & 0.92 (0.04) &            $-0.03$ (0.03) &           $-10.9$ (0.2) & Sbc \\
2001da          & 2001 Jul.\ 22.4                  &   4.8  (1.3) & 1.25 (0.05) & \phantom{$-$}0.16  (0.03) &           $-11.5$ (0.2) & Sab \\
2001el          & 2001 Sep.\ 24.8\tablenotemark{f} & $-5.3$ (0.1) & 1.15 (0.03) & \phantom{$-$}0.03  (0.03) &           $-11.7$ (0.2) & Scd \\
2007af          & 2007 Apr.\ 8.4                   &  24.2  (0.4) & 1.23 (0.05) & \phantom{$-$}0.00  (0.03) &           $-11.0$ (0.2) & Scd \\
2009le          & 2009 Nov.\ 26.0                  & $-4.8$ (0.1) & 0.96 (0.02) & \phantom{$-$}0.12  (0.02) &           $-12.1$ (0.2) & Sbc \\
2010A           & 2010 Jan.\ 8.0                   & $-8.2$ (0.1) & 0.88 (0.02) & \phantom{$-$}0.16  (0.02) &           $-10.3$ (0.2) & Sab \\
\hline
\multicolumn{6}{c}{Single/Symmetric} \\
\hline
2002cr          & 2002 May   13.2                  & $-1.3$ (0.5) & 1.23 (0.03) &            $-0.05$ (0.03) &           $-10.1$ (0.2) & Scd \\
2006cm          & 2006 Jun.\ 6.5                   &   7.0  (0.5) & 1.04 (0.07) & \phantom{$-$}1.07  (0.06) &           $-11.5$ (0.2) & Sb \\
2007sr          & 2008 Jan.\ 17.6                  &  33.3  (0.3) & 1.07 (0.07) & \phantom{$-$}0.09  (0.06) &           $-12.9$ (0.3) & Sm pec\\
\hline
\multicolumn{6}{c}{No Absorption} \\
\hline
2006eu          & 2006 Sep.\ 10.3                  &   2.3  (2.2) & 1.37 (0.14) & \phantom{$-$}0.49  (0.04) &           $-11.4$ (0.3) & E \\
2007hj          & 2007 Oct.\ 20.3                  &  43.8  (0.2) & 1.97 (0.06) & \phantom{$-$}0.12  (0.03) &           \nodata       & S0 \\
2007on          & 2008 Jan.\ 17.3                  &  62.0  (0.1) & 1.64 (0.01) & \phantom{$-$}0.12  (0.01) &           \nodata       & E \\
SNF20080514-002 & 2008 Jun.\ 13.4                  &  17.0  (0.5) & 1.39 (0.04) &            $-0.16$ (0.05) &           $-10.8$ (0.2) & S0 \\
2008hv          & 2008 Dec.\ 15.3                  & $-1.9$ (0.1) & 1.33 (0.03) & \phantom{$-$}0.03  (0.01) &           $-10.8$ (0.2) & S0 \\
2008ia          & 2008 Dec.\ 15.2                  &   2.0  (0.1) & 1.35 (0.01) & \phantom{$-$}0.00  (0.01) &           $-11.5$ (0.2) & S0 \\
2009nr          & 2010 Jan.\ 9.3                   &  12.5  (0.3) & 0.93 (0.03) & \phantom{$-$}0.00  (0.03) & \phantom{1}$-9.6$ (0.2) & Scd \\
\hline
\multicolumn{6}{c}{Ambiguous\tablenotemark{h}} \\
\hline
1994D           & 1994 Apr.\ 14\tablenotemark{i}   &  23.9  (0.4) & 1.42 (0.01) &            $-0.04$ (0.04) &           $-11.2$ (0.2) & S0\tablenotemark{j}

\enddata

\tablecomments{Uncertainties are listed in parentheses.}
\tablenotetext{a}{Days since $B$ maximum for high-resolution spectrum.}
\tablenotetext{b}{Exposure-time weighted mean observation time from \citet{DOdorico89}, which presented the highest S/N spectrum.  Observations span 1986 May 7.2 -- 11.1.}
\tablenotetext{c}{Exposure-time weighted mean observation time.  Observations span 2002 Oct.\ 30 -- Nov.\ 6.}
\tablenotetext{d}{Exposure-time weighted mean observation time.  Observations span 2002 Nov.\ 28 -- Dec.\ 2.}
\tablenotetext{e}{Date corresponding to the first epoch of spectroscopy from \citet{Patat07:06x}}
\tablenotetext{f}{Exposure-time weighted mean observation time.  Observations span 2001 Sep.\ 21.2 -- 28.3.}
\tablenotetext{g}{Exposure-time weighted mean observation time.  Observations span 2001 Oct.\ 16 -- 17.}
\tablenotetext{h}{See text for discussion of the classification of SN~1994D.}
\tablenotetext{i}{Date for \citet{Ho95} observation, which presented the highest S/N spectrum.}
\tablenotetext{j}{The host galaxy shows an obvious dust lane in images.}

\end{deluxetable*}

\subsection{Photometry}

\citetalias{Foley11:vgrad} presents 1630 \vsi\ measurements for 255
SNe~Ia \citep{Blondin12}.  Using a family of functions, they were able
to estimate \vsi\ at maximum brightness, \vsiz, for SNe~Ia with
moderate decline rates and \vsi\ measured within about a week of
maximum brightness.  This large sample provides the reference sample
for this work.  Although there are potential differences between this
sample and the underlying SN~Ia sample, the size of the sample
provides a reasonable baseline.  To further increase the overlap
between the expanded \citetalias{Sternberg11} sample and the
\citetalias{Foley11:vgrad} sample, here we provide new photometry,
spectroscopy, and/or derived values for seven SNe~Ia.

\citetalias{Foley11:vgrad} presented several light curve-derived SN
quantities ($t_{\rm max}$, $B_{\rm max} - V_{\rm max}$, and $\Delta
m_{15} (B)$) that are critical for measuring \vsiz.  The light curves
and original derived quantities were part of the CfA3
\citep{Hicken09:lc} and LOSS \citep{Ganeshalingam10} datasets.
\citet{Ganeshalingam10} also presented light-curve parameters for
SNF20080514-002, which we use for this study.  \citet{Foley12:09ig}
presented light-curve parameters for SN~2009ig, which we also use
here.  For SN~2009nr, we use the light-curve properties reported by
\citet{Khan11}.  Light curves of SNe~2007on, 2008C, 2008fp, 2008hv,
and 2008ia were presented in \citet{Stritzinger11}; we derive
light-curve parameters for these SNe by fitting the light curves with
the SNooPy Python package \citep{Burns11}.  We have corrected all
measurements of $B_{\rm max} - V_{\rm max}$ for Milky Way extinction.
All light-curve parameters are presented in Table~\ref{t:lowres}.

\begin{deluxetable*}{lllcrccc}
\tabletypesize{\scriptsize}
\tablewidth{0pt}
\tablecaption{Log of Low-Resolution Optical Spectral Observations and Basic Parameters\label{t:lowres}}
\tablehead{
\colhead{} &
\colhead{} &
\colhead{} &
\colhead{Time of} &
\colhead{} &
\colhead{$\Delta m_{15}(B)$} &
\colhead{\vsi} &
\colhead{\vsiz} \\
\colhead{SN} &
\colhead{UT Date} &
\colhead{Instrument} &
\colhead{Max.\ (MJD)} &
\colhead{Phase\tablenotemark{a}} &
\colhead{(mag)} &
\colhead{($10^{3}$ \kms)} &
\colhead{($10^{3}$ \kms)}}

\startdata

2007on                           & 2007 Nov.\ 17.2 & IMACS  & 54419.9 (0.1) &   1.3  (0.1) & 1.64 (0.01) &           $-10.9$ (0.1) &           \nodata () \\
2008C\tablenotemark{b}           & 2008 Jan.\ 5.3  & FAST   & 54466.1 (0.3) &   4.1  (0.3) & 1.07 (0.03) &           $-10.9$ (0.1) &           $-11.3$ (0.3) \\
SNF20080514-002\tablenotemark{c} & 2008 May   28.3 & FAST   & 54613.0 (0.5) &   1.3  (0.5) & 1.39 (0.04) &           $-10.7$ (0.1) &           $-10.8$ (0.2) \\
2008fp                           & 2008 Sep.\ 21.4 & WFCCD  & 54730.5 (0.1) & $-0.2$ (0.1) & 0.92 (0.01) &           $-10.7$ (0.1) &           $-10.7$ (0.2) \\
2008hv                           & 2008 Dec.\ 18.3 & LDSS-3 & 54817.1 (0.1) &   1.1  (0.1) & 1.33 (0.03) &           $-10.7$ (0.1) &           $-10.8$ (0.2) \\
2008ia                           & 2008 Dec.\ 15.3 & LDSS-3 & 54813.2 (0.1) &   2.1  (0.1) & 1.35 (0.01) &           $-11.3$ (0.1) &           $-11.5$ (0.2) \\
2009ig\tablenotemark{d}          & 2009 Sep.\ 5.4  & LRS    & 55080.0 (0.1) & $-0.4$ (0.1) & 0.89 (0.02) &           $-13.6$ (0.1) &           $-13.5$ (0.2) \\
2009le                           & 2009 Nov.\ 24.2 & FAST   & 55165.9 (0.1) & $-6.6$ (0.1) & 0.96 (0.02) &           $-12.8$ (0.1) &           $-12.1$ (0.2) \\
2009nr\tablenotemark{e}          & 2010 Jan.\ 7.5  & FAST   & 55192.7 (0.3) &  10.6  (0.3) & 0.93 (0.03) & \phantom{1}$-9.3$ (0.1) & \phantom{1}$-9.6$ (0.2) \\
2010A                            & 2010 Jan.\ 16.1 & FAST   & 55212.4 (0.1) & $-0.3$ (0.1) & 0.88 (0.02) &           $-10.3$ (0.1) &           $-10.3$ (0.2)

\enddata

\tablecomments{Uncertainties are listed in parentheses.}
\tablenotetext{a}{Days since $B$ maximum.}
\tablenotetext{b}{Velocity information originally presented by \citetalias{Foley11:vgrad}.}
\tablenotetext{c}{Light-curve parameters were originally reported by \citet{Ganeshalingam10}.}
\tablenotetext{d}{All measurements were originally reported by \citet{Foley12:09ig}.}
\tablenotetext{e}{Light-curve parameters were originally reported by \citet{Khan11}.}

\end{deluxetable*}

Additionally, SNe~2009le and 2010A were followed by the FLWO 1.2~m
telescope at Mt.\ Hopkins.  Their light curves will be published as
part of the CfA4 dataset (Hicken et~al., submitted), but their
light-curve parameters are presented here.  Again, the light curves
were also fit with SNOoPy, and their light-curve parameters are listed
in Table~\ref{t:lowres}.

\subsection{Low-Resolution Spectroscopy}

In addition to the new light-curve parameters listed above, we present
spectroscopic parameters here, specifically \vsi\ and \vsiz.  We use
the \citet{Foley12:09ig} spectroscopic measurements for SN~2009ig.
SN~2008C previously had sufficient light curves to determine a time of
maximum brightness \citepalias{Foley11:vgrad} but not a decline rate
or $B_{\rm max} - V_{\rm max}$.  With additional photometric data, we
were able to measure those quantities, but the measurement of \vsiz\
remains unchanged from that of \citetalias{Foley11:vgrad}.  For the
remaining SNe, we discuss the data acquisition and reduction below.

Additional low-resolution spectra were obtained with a variety of
instruments on several telescopes.  Specifically, we obtained data
using the FAST spectrograph \citep{Fabricant98} on the FLWO 1.5~m
telescope, the WFCCD spectrograph mounted on the Du~Pont 2.5~m
telescope, the IMACS spectrograph \citep{Dressler06} mounted on the
Magellan Baade 6.5~m telescope, and the LDSS-3 spectrograph
\citep{Allington-Smith94} mounted on the Magellan Clay 6.5~m
telescope.  Specifics for each spectrum are presented in
Table~\ref{t:lowres}.

Standard CCD processing and spectrum extraction of the low-resolution
data were accomplished with IRAF.  The data were extracted using the
optimal algorithm of \citet{Horne86}.  Low-order polynomial fits to
calibration-lamp spectra were used to establish the wavelength scale,
and small adjustments derived from night-sky lines in the object
frames were applied.  The IMACS and LDSS3 spectra were reduced with
IRAF routines as described by \citet{Hamuy06}.  For the remaining
spectra, we employed our own IDL routines to flux calibrate the data
and remove telluric lines using the well-exposed continua of
spectrophotometric standards \citep{Wade88, Foley03, Silverman12}.

All details of the low-resolution spectra are presented in
Table~\ref{t:lowres}.

We determine \vsi\ and \vsiz\ for these SNe using the same methods of
\citet{Blondin06} and \citetalias{Foley11:vgrad}.  The
\citetalias{Foley11:vgrad} method does not produce accurate \vsiz\
measurements for SNe~Ia with $\Delta m_{15} (B) > 1.5$~mag.  As a
result, we do not report \vsiz\ for SNe~Ia with $\Delta m_{15} (B) >
1.5$~mag.  Our measurements for \vsi\ and \vsiz\ are reported in
Table~\ref{t:lowres}.

We present the full cross-matched sample of SNe~Ia with both \vsiz\
measurements and high-resolution spectroscopy in
Table~\ref{t:highres}.  The SNe with both \vsiz\ measurements and
high-resolution spectroscopy are designated as the `F12' sample.


\section{Results}\label{s:results}

Using the F12 sample, we are able to compare different subsamples to
each other as well as to the F12 and \citetalias{Foley11:vgrad}
samples.  There are currently only four SNe~Ia in the Single/Symmetric
subsample (when including SN~1994D).  Although this subsample is
small, it is large enough to perform basic statistical analyses.
Nonetheless, we caution against over-interpretation of the results for
the Single/Symmetric subsample, and suggest that additional analysis
should be performed when the sample sizes increase.  For the results
presented here, we classify SN~1994D as Single; if SN~1994D is
classified instead as Blueshifted, the significance of our results is
reduced slightly, but not significantly.

\subsection{Phase}

The standard model to explain the variable Na~D seen in SNe~Ia
requires ionizing photons from the time of explosion (or immediately
after) to ionize the Na atoms in the circumstellar environment
\citep[e.g.,][]{Patat07:06x}.  After sufficient time, the Na atoms
would recombine, resulting in increased Na~D absorption.  If this
model is correct, one might worry that systems with outflows will on
average have no preferred line profile at early times (when the Na is
ionized), but will on average have blueshifted profiles at later
times.

To test this idea, we compared the different subsamples of the F12
sample.  The F12 sample has a median phase of 1.0~days relative to $B$
maximum.  The Blueshifted, Redshifted, Single/Symmetric, and No
Absorption subsamples have median phases of $-1.5$, $-2.3$, 15.5, and
12.5~days, respectively.  There is no indication that SNe in the
Blueshifted subsample tend to have later phases.

There are three possible scenarios that allow for the Blueshifted
subsample to predominantly come from progenitor systems with outflows
and not have a phase bias.  First, it is possible that the systems
with variable Na~D are well tuned such that the ionizing flux and the
distribution of the CSM are matched to show variability, while other
systems with outflows simply do not ionize the CSM because of a lack
of ionizing photons, the CSM being too far/dense, or both.
Alternatively, it may be the case that the recombination timescale for
most systems is shorter than the median time between explosion and the
time of the spectrum, or about two weeks (although this is unlikely
since variable Na~D is not seen until after maximum brightness even
when there are earlier observations).  Finally, progenitor systems may
have Na at various radii from the SN, where the closest Na is ionized,
but the more distant Na produces blueshifted absorption.  Indeed, the
two SNe~Ia with high-resolution spectroscopy and variable Na~D
(SNe~2006X and 2007le) have blueshifted Na~D absorption at all
spectroscopic epochs.

Regardless of why there is not an obvious phase bias in the F12
sample, it does not appear that the Blueshifted sample is missing a
significant number of potential outflow systems specifically because
the Na in the CSM was fully ionized at the time of spectroscopy.

\subsection{Light-Curve Shape}

We would like to compare the decline rate distributions of the various
samples.  The full high-resolution sample, including all SNe~Ia
regardless of decline rate (i.e., the F12 sample combined with SNe~Ia
with high-resolution spectroscopy and $\Delta m_{15} (B) > 1.5$~mag),
has a weighted mean $\Delta m_{15} (B)$ of $\mean{\Delta m_{15} (B)} = 1.22
\pm 0.01$~mag.  The Blueshifted, Redshifted, Single/Symmetric, and No
Absorption subsamples of the full high-resolution sample have weighted
means of $\mean{\Delta m_{15} (B)} = 1.21 \pm 0.01$, $1.10 \pm 0.02$,
$1.23 \pm 0.02$, and $1.45 \pm 0.03$~mag, respectively.  The subsample
of SNe~Ia with no Na~D absorption have significantly faster light
curves than the bulk of the sample.  Six of 7 SNe~Ia in this subsample
come from S0 or earlier host galaxies (Table~\ref{t:highres}), which
is not surprising since these host galaxies are expected to have
significantly less gas and dust, while no SN hosted in an elliptical
galaxy is in any of the other subsamples \citepalias{Sternberg11}.
SNe~Ia in elliptical galaxies tend to have significantly faster light
curves \citep[e.g.,][]{Howell01, Hicken09:de}.

The \citetalias{Foley11:vgrad} method of determining \vsiz\ does not
apply to SNe~Ia with $\Delta m_{15} (B) > 1.5$~mag.  As a result,
SNe~Ia with $\Delta m_{15} (B) > 1.5$~mag can not be included in the
F12 sample, and are therefore not included in any further analysis.
Since many of the SNe~Ia in the No Absorption subsample of the full
high-resolution sample have large decline rates, the F12 sample does
not include many SNe~Ia without Na~D absorption.  With this
restriction on light-curve shape, the F12 sample has a weighted mean
$\Delta m_{15} (B)$ of $\mean{\Delta m_{15} (B)} = 1.14 \pm 0.01$~mag,
and the Blueshifted, Redshifted, Single/Symmetric, and No Absorption
subsamples have weighted means of $\mean{\Delta m_{15} (B)} = 1.12 \pm
0.01$, $1.10 \pm 0.02$, $1.23 \pm 0.02$, and $1.25 \pm 0.02$~mag,
respectively.  The No Absorption subsample is still skewed to faster
declining SNe relative to the entire F12 sample.

\subsection{Maximum-Light Color}\label{ss:color}

We now compare the $B_{\rm max} - V_{\rm max}$ pseudo-color (after
correcting for Milky Way reddening) for the various samples.  The
\citetalias{Foley11:vgrad} sample has a median pseudo-color of
0.01~mag.  This is significantly bluer than the F12 sample, which has
a median pseudo-color of $B_{\rm max} - V_{\rm max} = 0.12$~mag.  A
K-S test results in a $p$-value of only 0.048, indicating that the two
samples may be drawn from the different parent population.

We interpret the color difference as a selection effect.  Since the SN
sample from which the \citetalias{Foley11:vgrad} sample is derived is
the sample of low-$z$ SNe discovered in SN surveys, there is some
amount of Malmquist bias associated with both the discovery and
follow-up of the sample.  This bias will naturally remove extinguished
SNe~Ia near the limit of SN searches from the sample.  However, for
high-resolution spectroscopy, SNe typically need to be much brighter
than the magnitude limit of SN searches; as a result, the Malmquist
bias in the F12 sample should be significantly less than the
\citetalias{Foley11:vgrad} sample.

The median pseudo-colors for the Blueshifted, Redshifted,
Single/Symmetric, and No Absorption subsamples are $B_{\rm max} -
V_{\rm max} = 0.34$, 0.08, 0.02, and 0.00~mag, respectively.  The
values for the Blueshifted, Single/Symmetric, and No Absorption
subsamples are fairly different from the median value for the full F12
sample.  However, the Single/Symmetric and No Absorption subsamples
have median pseudo-colors similar to that of the
\citetalias{Foley11:vgrad} sample, perhaps further indicating that
SNe~Ia in the Single/Symmetric and No Absorption subsamples have
minimal dust extinction.  Performing a K-S test between the
Blueshifted subsample and the F12 sample excluding the Blueshifted
subsample, we find a $p$-value of 0.021.  A similar test for the
Single/Symmetric and No Absorption subsamples results in $p$-values of
0.56 and 0.16, respectively.  Although the median values for all three
subsamples suggest that there might be differences in the underlying
populations of the subsamples, only the difference between the
Blueshifted subsample and the remainder of the F12 sample is
statistically significant.

The redder colors for the Blueshifted subsample are consistent with
there being some SNe~Ia in the sample with CSM resulting in additional
reddening that is combined with the underlying host-galaxy reddening
distribution.

\subsection{Ejecta Velocity}

\begin{figure}
\begin{center}
\epsscale{1.1}
\rotatebox{90}{
\plotone{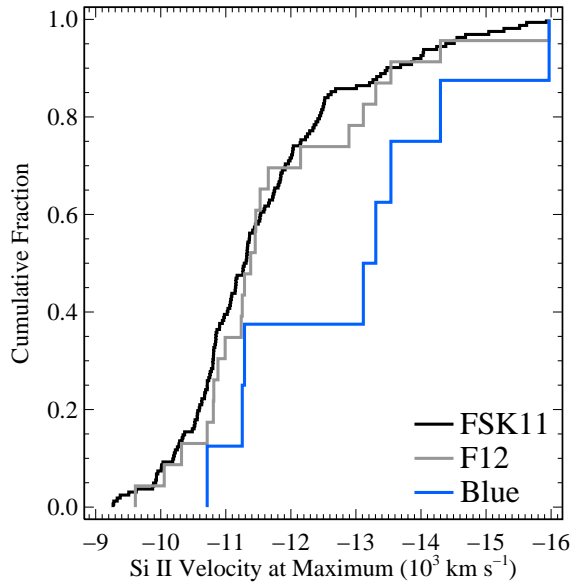}}
\caption{CDF of \vsiz\ for the \citetalias{Foley11:vgrad} sample
excluding SNe in the F12 sample (black line, labeled 'FSK11'), the F12
sample (gray line, labeled 'F12'), and the Blueshifted subsample (blue
line, labeled 'Blue').}\label{f:cdf}
\end{center}
\end{figure}

With the knowledge that Blueshifted and No Absorption SNe~Ia are
redder and faster decliners than the rest of the F12 sample,
respectively, while the other subsamples do not appear to be
significantly different from each other, we now examine correlations
between narrow Na~D profile properties and \vsiz.  Figure~\ref{f:cdf}
displays the cumulative distribution function (CDF) of \vsiz\ for
various samples.  The gray and black lines represent the F12 and
\citetalias{Foley11:vgrad} samples, respectively.  We also display the
Blueshifted subsample of the F12 sample with a blue line.

The CDF of the F12 sample has a very similar shape to that of the
\citetalias{Foley11:vgrad} sample.  The median \vsiz\ for both the
F12 sample and the \citetalias{Foley11:vgrad} samples is $-11$,300 and
$-11$,400~\kms, respectively.  Performing a K-S test (excluding any
SNe in the F12 sample from the \citetalias{Foley11:vgrad} sample), we
find a K-S $p$-value of 0.87.  From this test, there is no indication
that the F12 and \citetalias{Foley11:vgrad} samples have different
parent populations.

However, the Blueshifted subsample is qualitatively different from the
\citetalias{Foley11:vgrad} and F12 samples.  Interestingly, the five
SNe~Ia with the highest \vsiz\ in the F12 sample are all in the
Blueshifted subsample.  Specifically, 63\% of the Blueshifted
subsample has a higher \vsiz\ than {\it any} SN~Ia in any of the other
subsamples.  Not surprisingly, the CDF of the Blueshifted subsample
appears to be significantly different from that of the
\citetalias{Foley11:vgrad} sample or the F12 sample, with the
Blueshifted subsample tending to have higher \vsiz\ than the typical
SN~Ia.  The Blueshifted subsample has a median \vsiz\ of
$-13$,200~\kms, while the Redshifted, Single/Symmetric, and No
Absorption subsamples have median \vsiz\ of $-11$,300, $-11$,300, and
$-10$,800~\kms, respectively.  A K-S test to compare the Blueshifted
sample to the \citetalias{Foley11:vgrad} sample (again, excluding any
SNe~Ia from the F12 sample) results in a $p$-value of 0.034,
indicating that the two samples likely have different parent
distributions.

Although the F12 sample has a similar \vsiz\ distribution as the
\citetalias{Foley11:vgrad} sample, there may be other important ways
in which the parent populations of the two samples are different.  To
mitigate any potential bias, we performed K-S tests between the
Blueshifted subsample and the F12 sample (excluding the Blueshifted
subsamples).  This test resulted in a $p$-value of 0.018.  The
Blueshifted subsample appears to be drawn from a different parent
population than the complementary F12 sample.

We tested the robustness of this result using a Monte Carlo
simulation.  For each iteration, we selected a sample of 23 SNe (to
match the number of SNe~Ia for which we have high-resolution
spectroscopy) from the full \citetalias{Foley11:vgrad} sample of
SNe~Ia.  From this sample, we then selected a subsample of
``Blueshifted'' SNe equal to the number of Blueshifted SNe~Ia in the
F12 sample.  We performed a K-S test on these random samples similar
to that described above.  We also noted if the five highest or lowest
velocity SNe were in the ``Blueshifted'' sample.

We found that 1.1\% of all random samples had a K-S $p$-value as small
or smaller than that of the Blueshifted sample.  This value is similar
to the $p$-value that we found above (0.018), and indicates that the
$p$-value is a reasonable measurement of the probability of the
Blueshifted and F12 samples being randomly drawn from the same parent
distribution.  We also found that 0.28\% of the random samples had the
five highest or lowest velocity SNe in the ``Blueshifted'' subsample.
It is important to test both extremes since either case would have
been notable to us.  This test of the extreme SNe is perhaps a more
interesting statistic since it is highly plausible that some of the
Blueshifted systems come from the same parent population as the
Redshifted systems, but are simply confused through random
line-of-sight effects.  Furthermore, only 0.21\% of the random samples
had both of the above characteristics.  Although this last number is
something of an {\it a posteriori} measurement, it does indicate that
the number of extreme objects in the Blueshifted sample is providing
additional information beyond the simple K-S statistic.

\begin{figure}
\begin{center}
\epsscale{1.1}
\rotatebox{90}{
\plotone{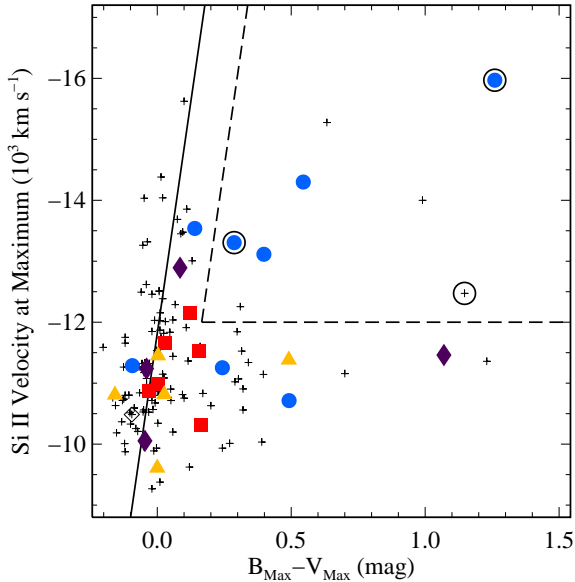}}
\caption{\vsiz\ as a function of $B_{\rm max} - V_{\rm max}$
pseudo-color for the \citetalias{Foley11:vgrad} (black crosses),
Blueshifted (blue circles), Redshifted (red squares), Single/Symmetric
(purple diamonds), and No Absorption subsamples (gold triangles).  The
solid line is the velocity-color relation between \vsiz\ and intrinsic
$B_{\rm max} - V_{\rm max}$ \citepalias{Foley11:vgrad}.  The empty
black circles indicate SNe~Ia with detected variable Na~D.  The region
defined by the dashed lines (corresponding to a 2-$\sigma$ deviation
from the \citetalias{Foley11:vgrad} relation between intrinsic $B_{\rm
max} - V_{\rm max}$ and \vsiz\ and $v_{\rm Si~II}^{0} = -12$,000~\kms)
has a large fraction of SNe~Ia with blueshifted and/or variable Na~D
line profiles.}\label{f:vel_bv}
\end{center}
\end{figure}

Compared to the F12 sample, the Blueshifted subsample appears to have
a statistically significant difference in \vsiz.  As seen in
Section~\ref{ss:color}, the SNe in the Blueshifted subsample also tend
to have redder observed $B_{\rm max} - V_{\rm max}$ colors.
\citet{Foley11:vel} and \citetalias{Foley11:vgrad} showed that ejecta
velocity and intrinsic color are highly correlated with SNe~Ia with
higher ejecta velocity tending to be intrinsically redder.  In
Figure~\ref{f:vel_bv}, we show \vsiz\ as a function of observed
$B_{\rm max} - V_{\rm max}$ for the \citetalias{Foley11:vgrad} and F12
samples.  We also display the \citetalias{Foley11:vgrad} relation
between intrinsic $B_{\rm max} - V_{\rm max}$ and \vsiz.  Several
SNe~Ia (including the majority of the Blueshifted subsample) fall far
off to the red of the \citetalias{Foley11:vgrad} relation, indicating
that their very red colors are dominated by dust reddening rather than
the trend with intrinsic color.

In the color-velocity parameter space, the Blueshifted subsample is
even more distinct from the rest of the F12 sample.  Performing a
two-dimensional K-S test on the Blueshifted subsample, the resulting
$p$-value of 0.023 is further indication of the Blueshifted subsample
being a distinct subpopulation in this parameter space.

In Figure~\ref{f:vel_bv}, we have also indicated the SNe~Ia for which
variable Na~D has been detected.  All three have high \vsiz\ and
$B_{\rm max} - V_{\rm max}$.  Two of those three have a blueshifted
Na~D profile, while the third does not have the high-resolution
spectroscopy necessary to make that measurement.  If one wants to
choose SNe~Ia which have a high probability of having variable Na~D,
one might want to choose reddened SNe~Ia with high ejecta velocities.
We defined a region of this parameter space corresponding to high
velocity ($v_{\rm Si~II}^{0} < -12$,000~\kms) and red (redder than the
2-$\sigma$ deviation from the \citetalias{Foley11:vgrad} relation
between intrinsic $B_{\rm max} - V_{\rm max}$ and \vsiz, which roughly
corresponds to $B_{\rm max} - V_{\rm max} > 0.2$~mag).  This region of
parameter space contains half of the Blueshifted subsample and all
SNe~Ia with observed Na~D variability.  We suggest that SNe~Ia in this
region are excellent candidates for having variable Na~D.
Additionally, special attention should be paid to SNe that have
Blueshifted Na~D absorption in the first epoch.


\section{Discussion \& Conclusions}\label{s:conc}

For the first time, we have shown a direct connection between the
progenitor system of a SN~Ia and its explosion properties.  In
contrast, previously, only indirect connections between progenitors
and SN~Ia properties have been made by examining host galaxy
properties \citep[e.g.,][]{Hamuy96:lum, Howell01, Hicken09:de}.
Specifically, we find that SNe~Ia with blueshifted narrow Na~D
profiles tend to have higher velocity ejecta as probed by \ion{Si}{2}
$\lambda 6355$ at maximum brightness than those with no Na~D
absorption or those with redshifted, single, or symmetric profiles.
This was achieved by expanding the \citetalias{Sternberg11} sample of
SNe~Ia with additional SNe~Ia with high-resolution spectroscopy and
matching that expanded high-resolution sample to the
\citetalias{Foley11:vgrad} sample of SNe~Ia with measured \vsiz,
again, expanding the \citetalias{Foley11:vgrad} sample slightly with
new data presented here.  The resulting sample is designated the `F12'
sample.

This result is further evidence that a significant fraction of SN~Ia
progenitor systems have outflows \citepalias{Sternberg11}, and that
the large fraction of SNe~Ia with blueshifted Na~D line profiles is
not caused by some other effect.  Similarly, this result supports the
interpretation that the variable Na~D is the result of ionized CSM
\citep{Patat07:06x, Simon09} rather than a geometric effect
\citep{Chugai08, Patat10}.

For most subsamples and the larger SN~Ia population, there is no
indication of different light-curve shape distributions.  However, the
No Absorption subsample tends to have faster declining light curves
(as expected given the host-galaxy distribution of the subsample).
There is no phase bias for the Blueshifted subsample, indicating that
few systems with outflows are artificially placed in other subsamples
because the circumstellar Na is completely ionized at the time of
spectroscopy.

The F12 sample has on average redder $B_{\rm max} - V_{\rm max}$
pseudo-colors than the \citetalias{Foley11:vgrad} sample.  We suggest
that this difference is the result of Malmquist bias reducing the
number of reddened SNe in the \citetalias{Foley11:vgrad} sample.
Compared to the rest of the F12 sample, the Blueshifted subsample is
slightly redder and the No Absorption subsample is slightly bluer.
The bluer color for the No Absorption subsample is consistent with a
SN population that has minimal dust reddening.  Similarly, the redder
color for the Blueshifted subsample is consistent with a SN population
with additional dust in the circumstellar environment.

High-resolution spectra of SNe~Ia in E/S0 galaxies have never shown
Na~D absorption (with the exception of SNe~Ia hosted in peculiar S0
galaxies, such SN~1986G, which was in Cen A).  Most single-degenerate
progenitor scenarios should exist in old stellar populations, so the
different host properties for the different subsamples indicate that
the population of SN progenitors with strong outflows may be extremely
rare in old stellar populations.  Alternatively, SNe~Ia from early and
late-type galaxies may have, on average, very different progenitor
systems.  Future studies of SN hosts and locations may constrain the
age of these progenitor systems, further constraining the exact
progenitor systems.

There are clear implications of the connection between blueshifted
Na~D absorption profiles and high ejecta velocities.  Specifically, if
all SNe~Ia come from a single progenitor channel, then one requires an
asymmetric progenitor system and explosion where the higher-velocity
ejecta are aligned with the higher density regions of the CSM or
explosions with higher kinetic energy per unit mass also have a denser
and/or closer CSM.  Alternatively, SNe~Ia could come from a variety of
progenitor channels where those that originate in progenitor systems
with strong outflows tend to have more kinetic energy per unit mass
than those with weak or no outflows.

From several different methods, specific progenitor systems which
should have strong outflows have been ruled out for SN~2011fe
\citep{Li11:11fe, Bloom12}, and there is strong evidence that its
progenitor system had a clean environment \citep{Chomiuk12, Patat12,
Margutti12}.  Since a significant fraction of SNe~Ia {\it do} have
strong progenitor outflows, it appears that there are at least two
progenitor channels for SNe~Ia.

\begin{acknowledgments} 

\bigskip
R.J.F.\ is supported by a Clay Fellowship.  A.G.Y\ is supported by
grants from the ISF and Minerva foundations, an ARCHES award and the
Lord Sieff of Brimpton Fund.  A.S.\ is supported by a Minerva
Fellowship.  M.H.\ acknowledges support by CONICYT through grants
FONDECYT Regular 1060808, Centro de Astrofisica FONDAP 15010003,
Centro BASAL CATA (PFB-06), and the Millenium Center for Supernova
Science (P10-064-F).  Supernova research at Harvard College
Observatory is supported by NSF grant AST--0907903.

We are grateful to S.\ Blondin and M.\ Hicken, who reduced some of the
data discussed in this work, and M.\ Shetrone, who assisted with the
HET data reduction..  We thank P.\ Challis, A.\ Filippenko, and H.\
Marion for looking for old data.  G.\ Folatelli helped organize the
observing at Las Campanas and M.\ Stritzinger observed one
low-resolution spectrum presented in this work.  We appreciate
comments from P.\ Nugent, F.\ Patat, L.\ Bildsten, A.\ Soderberg, and
L.\ Chomiuk.

We are grateful to the staffs at the Las Campanas, McDonald, and Fred
L.\ Whipple Observatories for their dedicated services.  The HET is a
joint project of the University of Texas at Austin, the Pennsylvania
State University, Stanford University,
Ludwig-Maximilians-Universit\"{a}t M\"{u}nchen, and
Georg-August-Universit\"{a}t G\"{o}ttingen. The HET is named in honor
of its principal benefactors, William P.\ Hobby and Robert E.\ Eberly.

\end{acknowledgments}

\bibliographystyle{fapj}
\bibliography{../astro_refs}


\end{document}